# Experiment of Multi-UAV Full-Duplex System Equipped with Directional Antennas

Tao Yu, Kento Kajiwara, Kiyomichi Araki, Kei Sakaguchi
Tokyo Institute of Technology, Japan
{yutao, kajiwara, araki, sakaguchi}@mobile.ee.titech.ac.jp

*Abstract*—One of the key enablers for the realization of a variety of unmanned aerial vehicle (UAV)-based systems is the high-performance communication system linking many UAVs and ground station. We have proposed a spectrum-efficient full-duplex directional-antennas-equipped multi-UAV communication system with low hardware complexity to address the issues of low spectrum efficiency caused by co-channel interference in areal channels. In this paper, by using the prototype system including UAVs and ground station, field experiments are carried out to confirm the feasibility and effectiveness of the proposed system's key feature, i.e., co-channel interference cancellation among UAVs by directional antennas and UAV relative position control, instead of energy-consuming dedicated self-interference cancellers on UAVs in traditional full-duplex systems. Both uplink and downlink performance are tested. Specially, in downlink experiment, channel power of interference between a pair of two UAVs is measured when UAVs are in different positional relationships. The experiment results agree well with the designs and confirm that the proposed system can greatly improve the system performance.

## I. INTRODUCTION

Due to high-performance control algorithms and downsizing of electronic components, terrestrial and aerial robots have become more functional, operable, productive, and inexpensive than ever. Autonomous mobility and unsupervised operation make robots appropriate unmanned platforms. The merits are especially amplified for systems employing multiple robots which can collaboratively execute complicated tasks, e.g., sensing, rescue, and exploration [1][2]. Multiple UAV-based systems especially gain our research interest because they further extend their mobility from 2-D to 3-D space and their working distances.

The high-performance wireless aerial communication system, which enables real-time information exchange between multiple UAVs and ground stations (GSs), is one of the key technologies to realizing multi-UAV systems and the related applications. However, there are still several issues hindering the development of high-performance multi-UAV aerial communication systems, such as follows. 1) Severe co-channel interference (CCI) among UAVs exists and results in low system performance and low spectrum efficiency, due to the lack of obstacles in the aerial propagation environment. 2) UAV-based systems have to take low energy consumption and low hardware complexity because of the limited battery-life and payload, but such constraints put a significant limit on the functionality and performance of functional modules on UAVs, including the communication system.

Most multi-UAV communication systems currently use existing standardized techniques, such as IEEE 802.11 families, cellular families, and even IEEE 802.15 families, or derivatives of them [3]-[5]. All of them are semi-duplex or out-band full-duplex (OBFD) systems. Namely, they cannot perform both transmission and reception with the same radio resources, which results in low radio resource efficiencies. Therefore, to solve the problem, the usage of in-band full-duplex (IBFD) communication systems is explored in UAV-based applications, such as full-duplex UAV relays [6]-[9]. Such systems typically suffer from severe self-interference and, consequently, the extra dedicated hardware and signal processing for the interference cancellations [10][11], which are big burdens and not suitable for UAVs. To address the issue, in our earlier papers [12][13], a multi-UAV IBFD (MU-IBFD) communication system with low hardware complexity was proposed for a multi-UAV-based video transmission and UAV control system. In the proposed system, both GS and UAVs are equipped with high-gain directional and beam steerable antennas to enable the systematic full-duplex communication for multiple UAVs.

Experimental research on aerial communication systems was also carried out, in addition to theoretical research. A testbed for a UAV-to-car communication system was constructed to evaluate the system performance with different UAV positions and antenna orientation/location [14]. In [15], a prototype of a UAV controlled over LTE link was reported, and the feasibility of the current LTE as UAV communication infrastructure was studied. In [16], field measurements for UAVs connected to LTE networks are conducted, and the performance of massive UAV deployment is analyzed by simulations. In [17], a mmWave-based UAV communication system is built for raw 4K video transmission. A comprehensive survey on prototypes and experiments of UAV communications can be found in [18].

In this paper, we conducted a field experiment in the air by using refitted UAVs equipped with developed prototype transceivers and directional antennas. The experiment is designed to validate the feasibility of the proposed MU-IBFD system and to show the key merit of the proposed architecture, i.e., CCI cancellation among UAVs by directional antennas and UAV position control instead of the energy-consuming dedicated cancellers, which are not suitable for UAVs.

The rest of this paper is organized as follows. Section II briefly recalls the system architecture. Section III describes the experimental system architecture and experimental procedure. The results are given and discussed in Section IV. Finally, Section V concludes this paper.

## II. SYSTEM ARCHITECTURE

To help readers understand the experiment, the system architecture of the MU-IBFD system is briefly recalled in this section. A more detailed explanation can be found in [12][13].

The topology of interest in this work is the star topology for multi-UAV applications such as remote sensing and wireless relay. Fig. 1(a) depicts the proposed MU-IBFD system. For simplicity, only two UAVs are shown in detail, because interference only occurs between two UAVs using

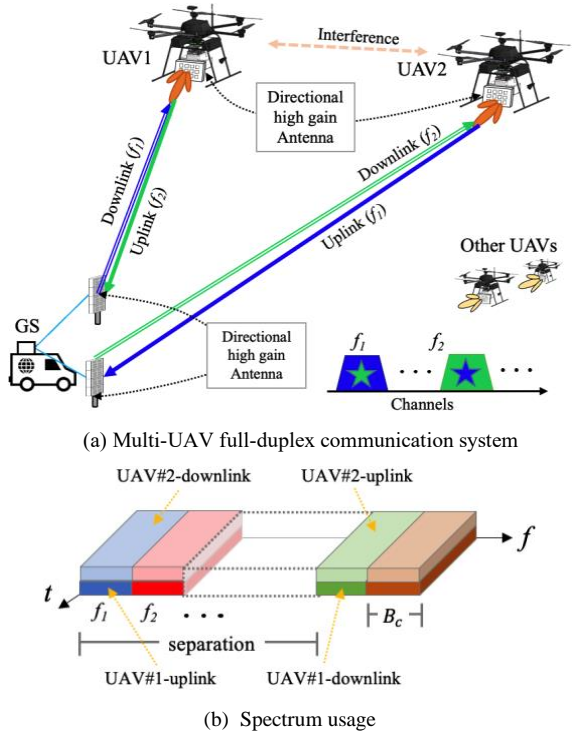

(a) Multi-UAV full-duplex communication system

(b) Spectrum usage

Fig. 1 Architecture of the proposed multi-UAV communication system

the same channels. Other pairs of UAVs follow the same mechanism. We mainly focus on interference on the UAV side and its influence on system performance, because GS does not have weight, dimension, or function limits as UAVs and can employ conventional IBFD design.

In the proposed system, the uplink and downlink of a UAV use two different and sufficiently separated channels, so complicated hardware for self-interference (SI) cancellation in conventional IBFD on UAVs can be avoided, and analog filters can effectively suppress adjacent-channel interference (ACI) between Tx/Rx. (In this study, uplink means UAV-to-GS link, and downlink means the opposite.) To improve spectrum efficiency, each channel is reassigned to uplink of a UAV and downlink of another. In Fig. 1(a), for example, UAV#1's uplink and UAV#2's downlink both use Channel#1 (blue); Channel#2 (green) is reused by UAV#1's downlink and uplink. All other UAV pairs reuse channels in the same manner. As a result, all channels are simultaneously used by the uplinks and downlinks in the system. The proposed duplex and multiplex schemes are depicted in Fig. 1(b). Therefore, from a system standpoint, IBFD communication is realized among multiple UAVs.

However, CCI between two UAVs becomes the key performance concern. The two strategies are proposed in the system to address the problem. 1) *Directional antennas*: All UAVs use high-gain directional beam-steerable antennas and beamforming, so high isolation between Tx and Rx of two UAVs using the same channel can be obtained passively in the propagation domain without further hardware and software cancellers. In addition, the high directional gain also improves signal power. 2) *UAV position control*: To further isolate Tx and Rx which reuse the same resource, the controllability, collaboration, and mobility of a multi-UAV system can be used. Based on antenna directivity patterns and positional relationships, UAVs can use active position control to limit CCI and ensure communication performance.

## III. DESCRIPTION OF EXPERIMENT

Experiments are designed and conducted to explore the feasibility and effectiveness of the proposed system. The experimental system, as shown in Fig. 2, includes two UAVs and a GS, all of which are fully functional. The channel assignments are as explained in Sec. II. Obviously, both of the two channels are re-used in two uplinks and downlinks, so it is an IBFD system from systematic standpoint.

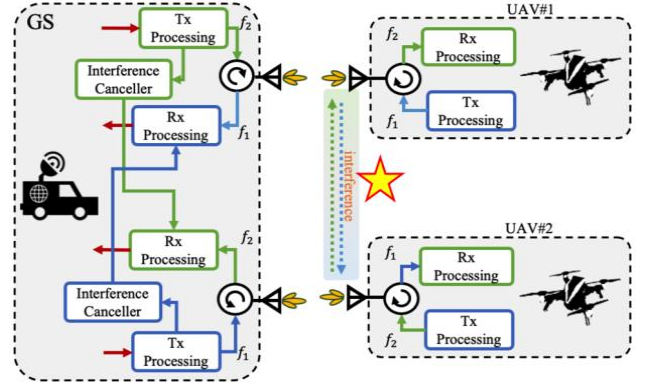

Fig. 2 Experimental system architecture

### A. Prototype UAVs and GS

The UAVs are refitted based on a DJI M600 Pro to carry the 4K camera, codecs, transceivers, dual-polarized directional antennas, antenna rotators, and other modules, as shown in Fig. 3. Mechanical beamforming is used by rotating the antenna mechanically according to the GS and UAV positions. GS is depicted in Fig. 4. A MU-IBFD long-distance (up to 5km) 4K(3840×2160)/60p video transmission (uplink) and control signal transmission (downlink) system is implemented as an example of UAV-based applications. 4K videos are compressed (to 40Mbps) and transmitted from UAVs to GS, and control signals are transmitted from GS to UAVs.

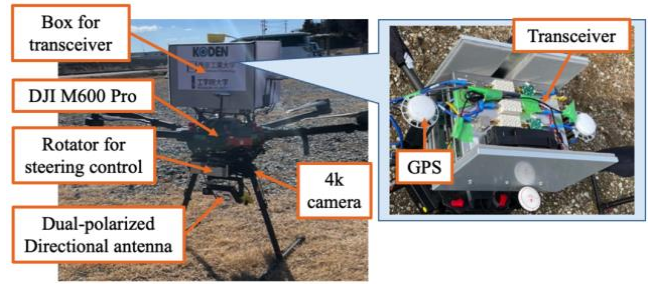

(a) Refitted UAV equipped with transceivers

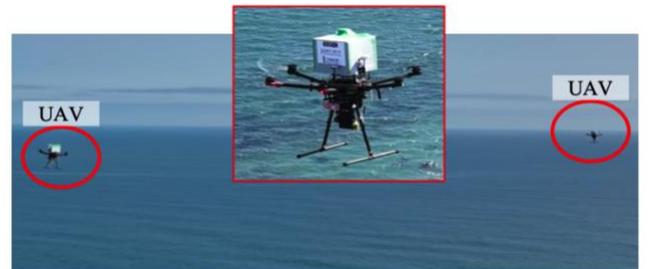

(b) Two UAVs in flight

Fig. 3 Prototype UAV

The physical layer design in the prototype system partially refers to the uplink in [19]. Single carrier modulation is adopted for higher energy efficiency. In downlink, spread spectrum is adopted to ensure the stability of transmission flight control. The prototype system works in 5.7 GHz band, and two channels with center frequencies of 5.725 GHz and 5.675 GHz are used by the two uplinks and downlinks. Each channel uses 10 MHz bandwidth (9 MHz transmission band, 1 MHz guard band). For a smaller size and higher communication capacity, polarization MIMO is used in UAVs, and each UAV is equipped with a dual-polarized directional patch antenna steered by a 2-axis rotator. No interference cancellers are used in UAVs, except for bandpass filters. Because there is no size and weight constraint in GS, it uses large and separated polarized slot antennas, as shown in Fig. 4. GS employs a conventional IBFD architecture, e.g., Tx-Rx isolations and analog cancellers. Very narrow and wide beamwidths in vertical and horizontal directions are designed for GS antennas so that GS uses fixed beam direction instead of dynamic beam tracking in the experiment. Metrics such as bit-error-rate (BER) and channel power can be measured and logged in transceivers to evaluate the communication performance. Detailed parameters are summarized in Table I.

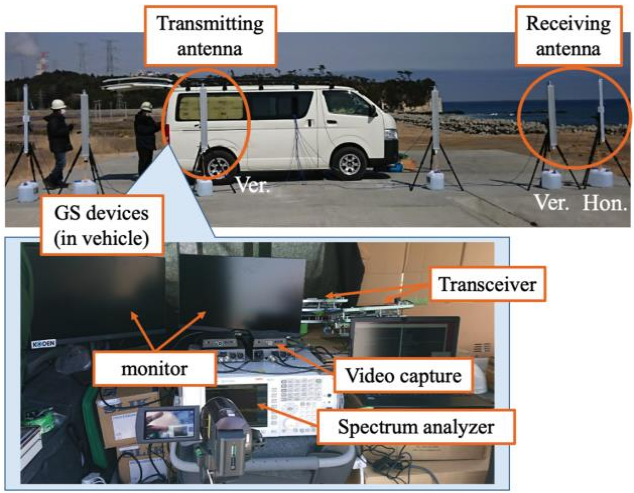

Fig. 4 Prototype GS

Table I Communication Parameters

| UAV | |
|---|---|
| Bandwidth | 9 MHz |
| Tx Power | 17 dBm / 17 dBm |
| Ant. Gain | 15 dBi |
| HPBW (ver./hon.) | 28 deg / 28 deg |
| Modulation | 16QAM-SC-FDM |
| Multiplexing | Polarization MIMO |
| GS | |
| Bandwidth | 9 MHz |
| Tx Power | 8.75 dBm |
| Ant. Gain (VP) | 18.2 dBi |
| HPBW (ver./hon.) | 3 deg / 52 deg |
| Ant. Gain (HP) | 19.8 dBi |
| HPBW (ver./hon.) | 3 deg / 75 deg |
| Modulation | QPSK-SS-SC-FDM |
| Multiplexing | SISO |

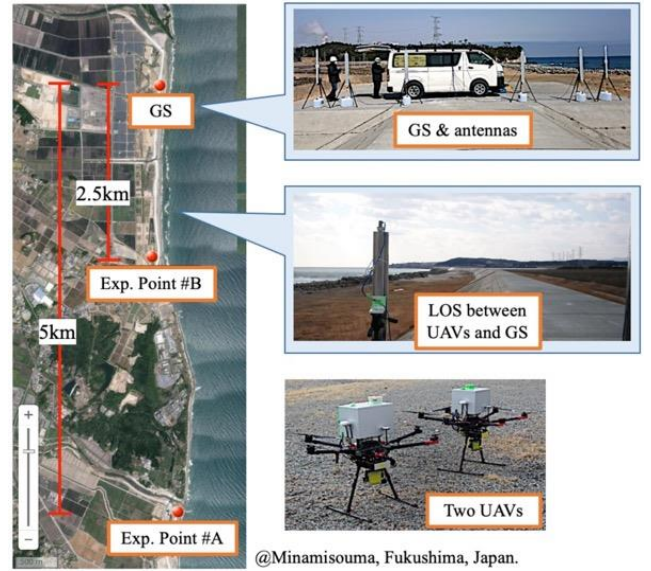

Fig. 5 Experiment environment

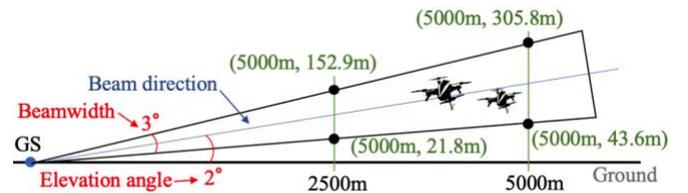

Fig. 6 Antenna Setup

### B. Multi-UAV Communication Experiment

For safety reasons and a long enough (up to 5 km) line-of-sight, the field experiment was conducted on the coast of Minamisoma, Fukushima, Japan. The experiment setup and environment are shown in Fig. 5. The GS and slot antennas were set up on a seawall which was built after the terrible earthquake and tsunami on March 11, 2011. LOS path aways existed between UAVs and GS. In the experiment, UAVs flew at around a height of 100m, and antennas were always directed towards the GS antenna by the rotator. A fixed GS antenna setup was employed due to the long GS-UAV distance, and the side view of the setup is shown in Fig. 6. UAVs were controlled to be within the half-power coverage of the GS antenna. Two experiments were conducted.

Uplink performance was tested near experiment point #A (around 5km away from GS). Two UAVs flew to a height of 100m and transmitted 4K videos to GS simultaneously. A photo of two UAVs in flight can be found in Fig. 3(b).

The downlink performance was tested near experiment point #B (about 2.5 km away from GS) because the terrain near point #B is convenient for UAV operations. In this paper, we focus more on the performance of downlink, which uses the proposed MU-IBFD architecture. The SINR and CCI were measured in detail when UAVs were in different relative positions. In experiment, UAV#2 hovered at a height of 100m and transmitted 4K video to GS via uplink, while UAV#1 was flying near UAV#2, as shown in Fig. 7(a). The CCI from UAV#2 to UAV#1 was measured and logged in UAV#1. During CCI measurement, the downlink of UAV#1, which uses the same channel as the uplink of UAV#2, was turned off, so that the received signal at UAV#1 was only CCI from UAV#2. Following CCI measurements, the uplink of UAV#2 was turned off, and the downlink of UAV#1 was turned on, allowing the received power of the expected signal to be measured, and the UAV#1 SINRs in different positions to be calculated. UAV#1 flew near UAV#2 along the path shown in Fig. 7(b) at different height (i.e., 100m, 90m, 80m, 70m). The paths and measurements are discrete and not uniform, so for ease of understanding, the data would be interpolated by a Gaussian process regression (GRP). Because the antenna patterns of two UAVs are the same and the CCI to each other is symmetric, only the influence from UAV#2 to UAV#1 was measured.

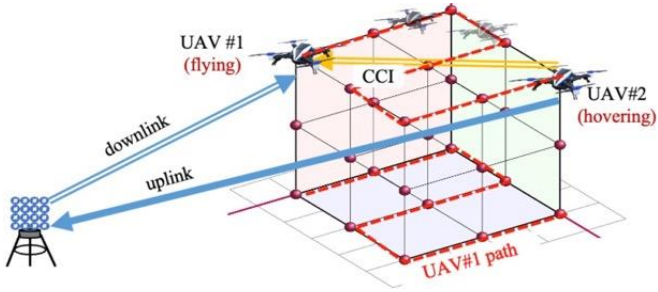

(a) Experiment scheme of CCI measurements

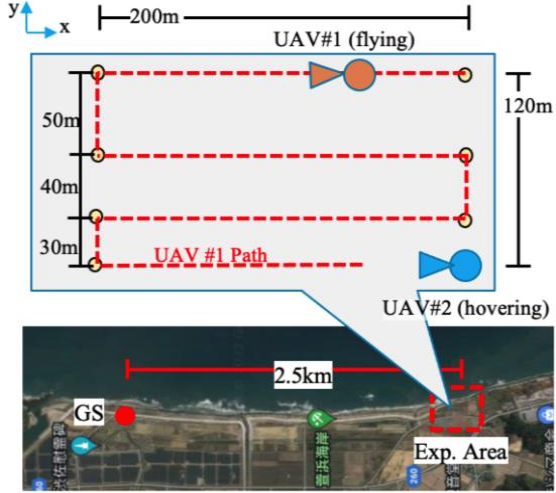

(b) UAV#1 flight path

Fig. 7 Experiment environment

## IV. EXPERIMENT RESULTS

### A. Uplink Experiment

Fig. 8(a) depicts monitors displaying 4K videos received at GS. It shows that 4K videos were successfully transmitted from UAVs 5km away to GS. Metrics including received signal strength indicators (RSSIs) (two streams) and BER of uplink signals were measured and logged by GS in the experiment. These metrics are shown as an example in Fig. 8(b) for 100 seconds of transmission in the experiment when UAVs were at a height of 100m and a distance of 5km. It is noted that because BER is calculated within one frame, $10^{-6}$ in the figure is considered as an error-free transmission.

Some transmission errors and RSSI drops occurred during the experiment, which can be observed in Fig. 8(b). They are considered to be mainly caused by beam errors due to the shaking of UAVs during flight, because most quadrotor UAVs, including the DJI M600Pro used in this experiment, have to frequently tilt the body when facing wind and turbulence during flight. To address such an issue, it is planned to introduce a 3-axis rotator and stabilizer for antenna steering control in the future.

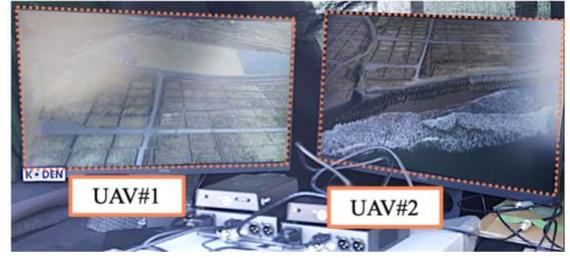

(a) Received video at GS

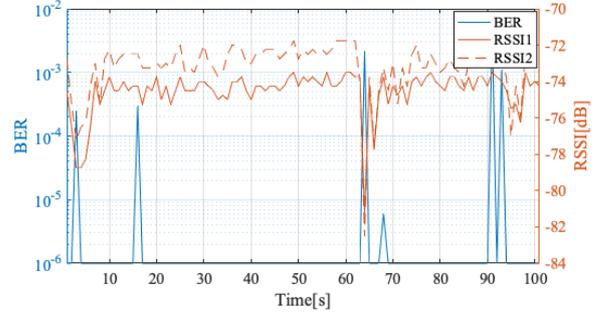

(b) BER and RSSIs at GS of one UAV link

Fig. 8 Uplink experiment results

### B. Downlink Experiment

As an example, Fig. 9(a) gives a series of CCI data collected at UAV#1 while flying along the path (from top right corner to bottom right corner) in Fig. 7(b) at the same height with UAV#2, i.e., 100m. The power of CCI and expected signal were measured at different heights of UAV#1 multiple times.

For ease of understanding, SINR is calculated and interpolated, as shown in Fig. 9(b). In the figure, UAV#2 is at (2500, 0, 100) and GS is at (0, 0, 1.5). Measurements were not taken in the area directly above UAV #2 for safety reasons as suggested by UAV operators, and this area has been left blank in Fig. 9(b). Fig. 9(b) confirms that SINR of UAV#1, affected by CCI, depends on the relative positions of two UAVs. Low SINR, i.e., high CCI, only occurs when two UAVs are close to each other and they are in the main lobe direction of each other. Otherwise, high SINR, i.e., low CCI, can be obtained. It is noted that the experiment area is very small compared to the practical working area and spacing separations of a group of UAVs. It indicates that with small effort to control the position of two UAVs, a comparable performance to the ideal IBFD systems can be achieved by the proposed MU-IBFD system. In Fig. 9(b), CCI in around 77% of the experiment area is lower than -90dBm. To achieve such a low level of interference in conventional IBFD systems, more than around 100 dB of SI needs to be mitigated by isolation and canceller, which is too hard to do in UAV-based communication systems.

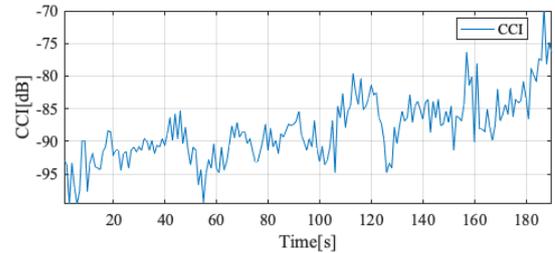

(a) Measured CCI of UAV#1 in 100m height

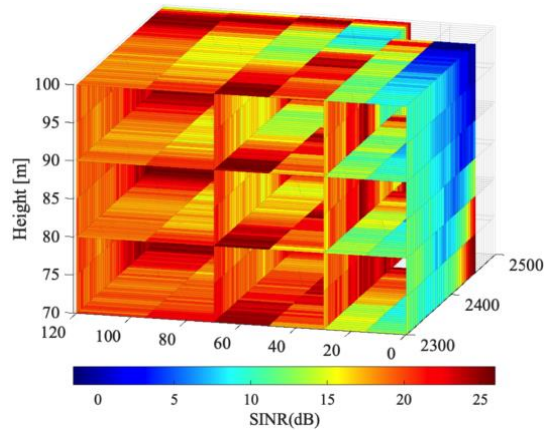

(b) Measured SINR of UAV#1

Fig. 9 Downlink experiment results

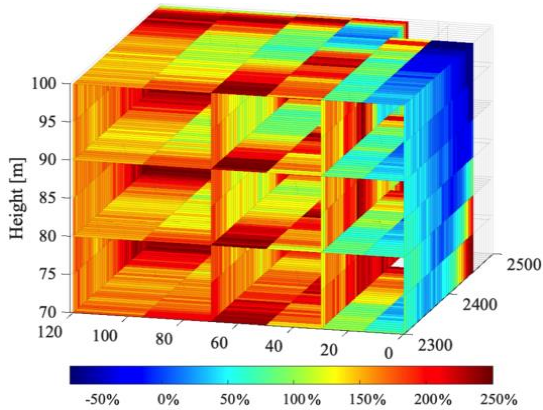

Fig. 10 Downlink capacity improvement compared to TDD/omni-directional antenna scheme in percentage

The downlink capacity of UAV#1 is also calculated by the measured SINRs in the experiment, and is compared with a conventional TDD/omni-directional antenna scheme with a 20% guard interval, a 36 dBm EIRP (maximum value allowed by local laws), and an ideal omni-directional antenna. Fig. 10 shows the capacity improvement in percentage, $\frac{C_{\text{MU-IBFD}} - C_{\text{TDD}}}{C_{\text{TDD}}} \times 100\%$, where $C_{\text{MU-IBFD}}$ and $C_{\text{TDD}}$ are the channel capacities of the MU-IBFD and TDD systems, respectively. The results show that compared to conventional TDD with an even much larger EIRP, performance drops of MU-IBFD only happen in the small area near the main lobe direction of UAV#2 (blue in Fig. 10). Otherwise, the downlink capacity of UAV#1 can be greatly increased by even much more than 100% thanks to the proposed MU-IBFD system.

## V. Conclusion

A field experiment is designed and conducted in this paper by using prototype hardware to confirm the feasibility of the proposed MU-IBFD system with low hardware complexity. The experiment result shows that CCI can be effectively mitigated by the proposed architecture. The consequent SINR and capacity highly depend on the relative positional relationship of the two UAVs. It shows that much higher system performance can be achieved if the positions of two UAVs are well controlled. The results also demonstrate the superior performance of MU-IBFD compared with the conventional TDD scheme, even with a much higher EIRP.

## Acknowledgment

This work was supported by Japanese Ministry of Internal Affairs and Communications (MIC) under the grant agreement 0155-0083.